\documentclass[twocolumn]{aastex631}
\usepackage{graphicx}

\newcommand{\tnm}[1]{\tablenotemark{#1}}

\newcommand{\kms}{\mbox{km\ s$^{-1}$}}
\newcommand{\kmsMpc}{\kms~\mbox{Mpc}$^{-1}$}
\newcommand{\MUV}{\ensuremath{M_{UV}}}
\newcommand{\Msun}{\ensuremath{M_{\odot}}}

\newcommand{\Zsun}{\ensuremath{Z_{\odot}}}

\newcommand{\Mstar}{\ensuremath{M_{*}}}

\newcommand{\logmstar}{$\log M_{\star}/\Msun$}
\newcommand{\sfr}{$\Msun/{\rm yr}$}

\newcommand{\mumean}{\ensuremath{\mu_{\rm mean}}}
\newcommand{\LCDM}{$\Lambda$CDM}
\newcommand{\Ho}{\ensuremath{H_{0}}}
\newcommand{\Om}{\Omega_{\rm m}}
\newcommand{\OL}{\Omega_{\Lambda}}
\newcommand{\dg}{\ensuremath{^\circ}}
\newcommand{\mAB}{\ensuremath{m_{\rm AB}}}
\newcommand{\zphot}{\ensuremath{z_{\rm phot}}}
\newcommand{\zspec}{\ensuremath{z_{\rm spec}}}
\newcommand{\zbest}{\ensuremath{z_{\rm best}}}
\newcommand{\zform}{\ensuremath{z_{\rm form}}}
\newcommand{\tform}{\ensuremath{t_{\rm form}}}

\newcommand{\fnu}{\ensuremath{f_{\nu}}}
\newcommand{\Pz}{\ensuremath{{\rm P}(z)}}
\newcommand{\Pzgtr}[1]{\ensuremath{{\rm P}(z > #1)}}
\newcommand{\dz}{\ensuremath{{\rm d}z}}
\newcommand{\HST}{\textit{HST}}

\newcommand{\JWST}{\textit{JWST}}
\newcommand{\Hubble}{\textit{Hubble Space Telescope}}

\newcommand{\Webb}{\textit{James Webb Space Telescope}}

\newcommand{\wl}{\mbox{$\lambda$}}
\newcommand{\ionrm}[1]{\mbox{\small\sc{\romannumeral #1}}}
\newcommand{\forb}[2]{\mbox{[#1~\ionrm{#2}]}}

\newcommand{\forbww}[4]{\mbox{[#1~\ionrm{#2}]~\wl\wl#3,~#4}}

\newcommand{\OIIIww}{\forbww{O}{3}{4959}{5007}}
\newcommand{\OIII}{\forb{O}{3}}

\newcommand{\Hb}{\mbox{H\ensuremath{\beta}}}
\newcommand{\HII}{H\,\textsc{ii}}
\newcommand{\Lya}{Lyman-\ensuremath{\alpha}}

\newcommand{\WHL}{WHL0137$-$08}
\newcommand{\idnum}[1]{WHL0137--ID#1}
\newcommand{\arcid}{\idnum{13362}}

\newcommand{\grizli}{\textsc{grizli}}
\newcommand{\eazypy}{\textsc{eazypy}}
\newcommand{\bagpipes}{\textsc{bagpipes}}
\newcommand{\beagle}{\textsc{beagle}}
\newcommand{\photutils}{\textsc{photutils}}

\newcommand{\astropy}{\textsc{astropy}}
\newcommand{\astrodrizzle}{\textsc{astrodrizzle}}
\newcommand{\multinest}{\textsc{multinest}}
\newcommand{\cloudy}{\textsc{Cloudy}}
\newcommand{\jdaviz}{\textsc{Jdaviz}}

\submitjournal{ApJ}

\begin{document}

\title{High-Redshift Galaxy Candidates at $z = 9 - 10$ as Revealed by \JWST\ Observations of WHL0137-08}

\correspondingauthor{Larry D. Bradley}
\email{lbradley@stsci.edu}

\author[0000-0002-7908-9284]{Larry D. Bradley}
\affiliation{Space Telescope Science Institute (STScI), 3700 San Martin Drive, Baltimore, MD 21218, USA}

\author[0000-0001-7410-7669]{Dan Coe}
\affiliation{Space Telescope Science Institute (STScI), 3700 San Martin Drive, Baltimore, MD 21218, USA}
\affiliation{Association of Universities for Research in Astronomy (AURA) for the European Space Agency (ESA), STScI, Baltimore, MD, USA}
\affiliation{Center for Astrophysical Sciences, Department of Physics and Astronomy, The Johns Hopkins University, 3400 N Charles St. Baltimore, MD 21218, USA}

\author[0000-0003-2680-005X]{Gabriel Brammer}
\affiliation{Cosmic Dawn Center (DAWN), Copenhagen, Denmark}
\affiliation{Niels Bohr Institute, University of Copenhagen, Jagtvej 128, Copenhagen, Denmark}

\author[0000-0001-6278-032X]{Lukas J. Furtak}
\affiliation{Physics Department, Ben-Gurion University of the Negev, P.O. Box 653, Be'er-Sheva 84105, Israel}

\author[0000-0003-2366-8858]{Rebecca L. Larson}
\altaffiliation{NSF Graduate Fellow}
\affiliation{The University of Texas at Austin, Department of Astronomy, Austin, TX, United States}

\author[0000-0002-5588-9156]{Vasily Kokorev}
\affiliation{Kapteyn Astronomical Institute, University of Groningen, P.O. Box 800, 9700AV Groningen, The Netherlands}

\author[0000-0002-8144-9285]{Felipe Andrade-Santos}
\affiliation{Department of Liberal Arts and Sciences, Berklee College of Music, 7 Haviland Street, Boston, MA 02215, USA}
\affiliation{Center for Astrophysics \text{\textbar} Harvard \& Smithsonian, 60 Garden Street, Cambridge, MA 02138, USA}

\author[0000-0003-0883-2226]{Rachana Bhatawdekar}
\affiliation{European Space Agency, ESA/ESTEC, Keplerlaan 1, 2201 AZ Noordwijk, NL}

\author[0000-0001-5984-0395]{Maru{\v s}a Brada{\v c}}
\affiliation{University of Ljubljana, Department of Mathematics and Physics, Jadranska ulica 19, SI-1000 Ljubljana, Slovenia}
\affiliation{Department of Physics and Astronomy, University of California Davis, 1 Shields Avenue, Davis, CA 95616, USA}

\author[0000-0002-8785-8979]{Tom Broadhurst}
\affiliation{Department of Theoretical Physics, University of the Basque Country UPV/EHU, Bilbao, Spain}
\affiliation{Donostia International Physics Center (DIPC), 20018 Donostia,Spain}
\affiliation{IKERBASQUE, Basque Foundation for Science, Bilbao, Spain}

\author[0000-0002-1482-5818]{Adam Carnall}
\affiliation{Institute for Astronomy, University of Edinburgh, Royal Observatory, Edinburgh EH9 3HJ, UK}

\author[0000-0003-1949-7638]{Christopher J. Conselice}
\affiliation{Jodrell Bank Centre for Astrophysics, University of Manchester, Oxford Road, Manchester UK}

\author[0000-0001-9065-3926]{Jose M. Diego}
\affiliation{Instituto de F\'isica de Cantabria (CSIC-UC). Avda. Los Castros s/n. 39005 Santander, Spain}

\author[0000-0003-1625-8009]{Brenda Frye}
\affiliation{Department of Astronomy, Steward Observatory, University of Arizona, 933 North Cherry Avenue, Tucson, AZ 85721, USA}

\author[0000-0001-7201-5066]{Seiji Fujimoto}\altaffiliation{Hubble Fellow}
\affiliation{
Department of Astronomy, The University of Texas at Austin, Austin, TX 78712, USA
}
\affiliation{Cosmic Dawn Center (DAWN), Copenhagen, Denmark}
\affiliation{Niels Bohr Institute, University of Copenhagen, Jagtvej 128, Copenhagen, Denmark}

\author[0000-0003-4512-8705]{Tiger Y.-Y Hsiao}
\affiliation{Center for Astrophysical Sciences, Department of Physics and Astronomy, The Johns Hopkins University, 3400 N Charles St. Baltimore, MD 21218, USA}

\author[0000-0001-6251-4988]{Taylor A. Hutchison}
\altaffiliation{NASA Postdoctoral Fellow}
\affiliation{Astrophysics Science Division, NASA Goddard Space Flight Center, 8800 Greenbelt Rd, Greenbelt, MD 20771, USA}

\author[0000-0003-1187-4240]{Intae Jung}
\affiliation{Space Telescope Science Institute (STScI), 3700 San Martin Drive, Baltimore, MD 21218, USA}

\author[0000-0003-3266-2001]{Guillaume Mahler}
\affiliation{Institute for Computational Cosmology, Durham University, South Road, Durham DH1 3LE, UK}
\affiliation{Centre for Extragalactic Astronomy, Durham University, South Road, Durham DH1 3LE, UK}

\author[0000-0003-0503-4667]{Stephan McCandliss}
\affiliation{Center for Astrophysical Sciences, Department of Physics and Astronomy, The Johns Hopkins University, 3400 N Charles St. Baltimore, MD 21218, USA}

\author[0000-0003-3484-399X]{Masamune Oguri}
\affiliation{Center for Frontier Science, Chiba University, 1-33 Yayoi-cho, Inage-ku, Chiba 263-8522, Japan}
\affiliation{Department of Physics, Graduate School of Science, Chiba University, 1-33 Yayoi-Cho, Inage-Ku, Chiba 263-8522, Japan}

\author[0000-0002-9365-7989]{Marc Postman}
\affiliation{Space Telescope Science Institute (STScI), 3700 San Martin Drive, Baltimore, MD 21218, USA}

\author[0000-0002-7559-0864]{Keren Sharon}
\affiliation{Department of Astronomy, University of Michigan, 1085 S. University Ave, Ann Arbor, MI 48109, USA}

\author[0000-0001-9391-305X]{M. Trenti}
\affiliation{School of Physics, University of Melbourne, Parkville 3010, VIC, Australia}
\affiliation{ARC Centre of Excellence for All Sky Astrophysics in 3 Dimensions (ASTRO 3D), Australia}

\author[0000-0002-5057-135X]{Eros Vanzella}
\affiliation{INAF -- OAS, Osservatorio di Astrofisica e Scienza dello Spazio di Bologna, via Gobetti 93/3, I-40129 Bologna, Italy}

\author[0000-0003-1815-0114]{Brian Welch}
\affiliation{Department of Astronomy, University of Maryland, College Park, MD 20742, USA}
\affiliation{Observational Cosmology Lab, NASA Goddard Space Flight Center, Greenbelt, MD 20771, USA}
\affiliation{Center for Research and Exploration in Space Science and Technology, NASA/GSFC, Greenbelt, MD 20771}

\author[0000-0001-8156-6281]{Rogier A. Windhorst}
\affiliation{School of Earth and Space Exploration, Arizona State University,
Tempe, AZ 85287-1404, USA}

\author[0000-0002-0350-4488]{Adi Zitrin}
\affiliation{Physics Department, Ben-Gurion University of the Negev, P.O. Box 653, Be'er-Sheva 84105, Israel}



\begin{abstract}

We report the discovery of four galaxy candidates observed 450 -- 600
Myr after the Big Bang with photometric redshifts between $z \sim 8.3 -
10.2$ measured using \Webb\ (\JWST) NIRCam imaging of the galaxy cluster
\WHL\ observed in 8 filters spanning 0.8--5.0 $\mu$m, plus 9 \Hubble\
filters spanning 0.4--1.7 $\mu$m. One candidate is gravitationally
lensed with a magnification of $\mu \sim 8$, while the other three
are located in a nearby NIRCam module with expected magnifications of
$\mu \lesssim 1.1$. Using SED fitting, we estimate the stellar masses
of these galaxies are typically in the range \logmstar = $8.3 - 8.7$.
All appear young with mass-weighted ages $< 240$ Myr, low dust content
$A_V < 0.15$ mag, and specific star formation rates sSFR $\sim 0.25 -
10$ Gyr$^{-1}$ for most. One $z \sim 9$ candidate is consistent with
an age $< 5$ Myr and a sSFR $\sim 10$ Gyr$^{-1}$, as inferred from a
strong F444W excess, implying \OIII+\Hb\ rest-frame equivalent width
$\sim$2000 \AA, although an older $z \sim 10$ object is also allowed.
Another $z\sim 9$ candidate is lensed into an arc 2\farcs4 long with a
magnification of $\mu \sim 8$. This arc is the most spatially-resolved
galaxy at $z \sim 9$ known to date, revealing structures $\sim$30
pc across. Follow-up spectroscopy of \WHL\ with \JWST/NIRSpec will
be useful to spectroscopically confirm these high-redshift galaxy
candidates and to study their physical properties in more detail.

\end{abstract}



\section{Introduction} \label{sec:intro}

The \Webb\ (\JWST), with its 6.5m aperture and infrared capabilities
\citep{Rigby2023}, has opened a new window to study galaxies
in the early universe. In the first weeks of \JWST\ science
observation, a wealth of distant galaxy candidates \citep{Naidu2022a,
Castellano2022, Donnan2023, Finkelstein2022, Adams2023, Atek2023,
Harikane2023} were reported from the \JWST\ Early Release Observations
\citep[ERO;][]{Pontoppidan2022} and the Cosmic Evolution Early Release
Science (CEERS) \citep{Finkelstein2017} and Through the Looking GLASS
(GLASS-JWST) \citep{Treu2022} Early Release Science (ERS) programs
that surpass the distance record set by the \Hubble\ (\HST) at
$z=11.1$ \citep{Oesch2016}. These independent studies have revealed an
unexpectedly large abundance of bright galaxies ($\MUV \lesssim -21$;
e.g., \citealt{Finkelstein2022, Naidu2022a, Atek2023, Furtak2023})
that could pose a challenge to our current models of galaxy formation
\citep{Naidu2022b, Ferrara2023, Harikane2023}. Similarly, some $z \sim
7 - 11$ candidates were reported to have surprisingly large stellar
masses $\Mstar > 10^{10} \Msun$ \citep{Labbe2023} in apparent tension
with $\Lambda$CDM \citep{Boylan-Kolchin2023, Lovell2023} unless these
galaxies have lower masses \citep{Endsley2023, Steinhardt2023} or
incorrect redshifts.

Simulations suggest we should not have expected to find overly
massive galaxies in early \JWST\ observations, but rather that
we have likely only discovered the youngest, most actively star
forming galaxies given imaging depths to date of AB mag $\sim$29
\citep{Mason2023}. Analyses of these $z \sim 9 - 16$ candidates
observed in \JWST\ imaging further reveal young stellar ages $\sim~10
- 100$ Myr \citep{Whitler2023,Furtak2023}, younger than the median
ages $\sim 100$ Myr measured at slightly lower redshifts $z \sim
7 - 9$ \citep{Leethochawalit2023,Endsley2023}. Evidence that some
of these galaxies are extremely young, $< 10$ Myr, $z \sim 7 - 9$
is provided by very strong emission lines in NIRSpec spectroscopy
\citep{Carnall2023,Tacchella2023,Trussler2023}, with flux excesses also
clearly observed in photometry, especially when imaging is available in
four NIRCam long-wavelength filters F277W, F356W, F410M, and F444W.

Gravitational lensing by massive galaxy clusters can address these
problems in some detail as it provides magnified distant galaxies,
boosting their luminosity and revealing small-scale structures that
would otherwise be unobservable. Using these ``cosmic telescopes",
surveys such as CLASH \citep{Postman2012}, the Hubble Frontier Fields
\citep{Lotz2017}, and RELICS \citep{Coe2019} have revealed hundreds
of galaxy candidates in the reionization epoch. Using this technique,
we have discovered highly magnified \citep{Bradley2008, Zheng2012,
Hoag2017, Bouwens2014, Bradley2014, Infante2015, Salmon2018, Salmon2020}
and multiply imaged galaxies \citep{Frye1998, Frye2008, Coe2013,
Zitrin2014} at redshifts up to $z \sim 10.8$, many of which were the
most-distant known galaxy at the time of their discovery. Recent
\JWST\ observations of lensing clusters have pushed this frontier even
further with many high-redshift candidates detected in ERO observations
of the massive galaxy cluster SMACS0723 \citep{Adams2023, Atek2023,
Morishita2023, Donnan2023, Harikane2023} and the GLASS-JWST cluster
Abell~2744 \citep{Naidu2022a, Castellano2022, Leethochawalit2023,
Donnan2023, Harikane2023}, with photometric redshifts out to $z \sim 16$
\citep{Atek2023}.

Gravitationally lensed galaxies have allowed us to place
strong constraints on the evolution of the galaxy ultraviolet
luminosity function and the star formation rate density at $z > 8$
\citep{Bouwens2014, Atek2015, Ishigaki2015, Laporte2015, Livermore2017,
Bouwens2017, Atek2018, Bhatawdekar2019, Bouwens2022}. Gravitational
lensing has also provided us the ability to study small-scale structures
and star clusters within high-redshift galaxies down to scales of a few
parsec \citep[e.g.,][]{Vanzella2022, Mestric2022, Welch23_clumps}.

The Reionization Lensing Cluster Survey (RELICS) \HST\ Treasury Program
\citep{Coe2019} was designed to efficiently discover high-redshift
galaxy candidates bright enough for follow-up observations with
current and future observatories, including the Atacama Large
Millimeter/submillimeter Array (ALMA) and \JWST. By observing 41
strong lensing galaxy clusters with {\em Hubble} and {\em Spitzer},
RELICS discovered and studied over 300 high-redshift candidates in
the first billion years \citep{Salmon2020, Strait2021}, including the
brightest robust candidates known at $z \sim 6$, the Sunrise Arc,
a 2\farcs5 long arc at $z \sim 6$ \citep{Salmon2020}, and the most
distant spatially-resolved lensed arc, SPT0615-JD1, at $z \sim 10$
\citep{Salmon2018}. Remarkably, the RELICS survey also discovered the
gravitationally lensed star WHL0137-LS, nicknamed Earendel, with a
photometric redshift $\zphot = 6.2 \pm 0.1$ \citep{Welch2022_earendel}.

Earendel was discovered within the $z \sim 6$ Sunrise Arc
\cite{Welch23_clumps} lensed by the massive galaxy cluster WHL
J013719.8--082841 (hereafter \WHL; RA = 01:37:25.0, Dec = $-$08:27:23,
J2000), which is the focus of this paper. \WHL\ was discovered
by \cite{Wen2012} based on photometric redshifts in SDSS-III DR8
\citep{SDSS3DR8} and has a spectroscopic redshift of $z = 0.566$ based
on two cluster members within $r_{500} = 0.82$~Mpc from its brightest
cluster galaxy \citep{Wen2015}. The Planck SZ survey also identified
this cluster (WHL-J24.3324-8.477) as the 31st most massive in the Planck
PSZ2 catalog with $M_{500} = (8.9 \pm 0.7) \times 10^{14} M_{\odot}$
\citep{PSZ2}.

In this paper, we present high-redshift candidates at $z \sim 8.3 -
10.2$ discovered in \JWST\ NIRCam imaging of \WHL, obtained primarily to
study Earendel \citep{Welch2022_earendel} and the Sunrise Arc in more
detail. Our sample includes both a strongly-lensed galaxy candidate
discovered behind the cluster and three candidates in the nearby NIRCam
module, centered $\sim$ 2\farcm9 from the cluster center, with weak
magnifications of $\mu \le 1.1$. We use the AB magnitude system, $\mAB
= 31.4 - 2.5 \log(\fnu\ / {\rm nJy})$ \citep{Oke1974,OkeGunn1983}.
Where needed, we adopt a {\em Planck} 2018 flat \LCDM\ cosmology
\citep{Planck18_cosmo} with $\Ho = 67.7$~\kmsMpc, $\Om = 0.31$, and $\OL
= 0.69$, for which the universe is 13.8 billion years old and $1\arcsec
\sim 4.6$ kpc at $z = 9$.

All of the \JWST\ and \HST\ data of \WHL\ are public. Reduced images,
catalogs, lens models, and analysis code are available via our
website.\footnote{\url{https://cosmic-spring.github.io}}

\begin{figure*}[pt!]
\plotone{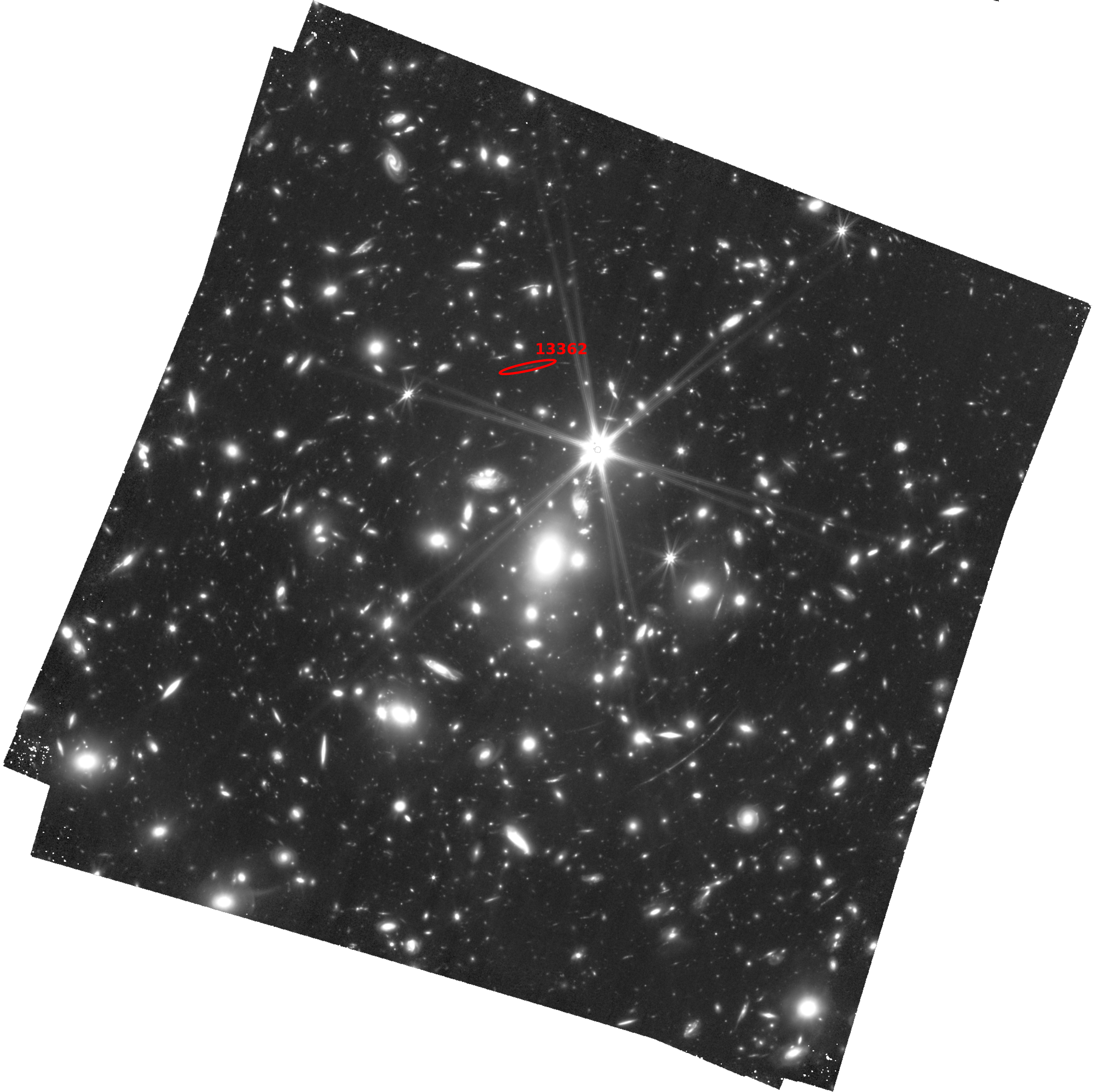}

\caption{
\JWST\ NIRCam detection image, comprised of a weighted sum of all NIRCam
LW images (F277W, F356W, F410M, F444W), of the \WHL\ cluster field. The
field of view is $\sim 2\farcm3 \times 2\farcm3$ and the image is shown
with north up and east left. The location of the strongly lensed $z \sim
9$ high-redshift candidate in the \WHL\ cluster field is indicated with
a red ellipse. The locations of the other high-redshift candidates are
shown in Figure~\ref{fig:nircam_par}.
\label{fig:nircam_cluster}}
\end{figure*}

\begin{figure*}[pt!]
\plotone{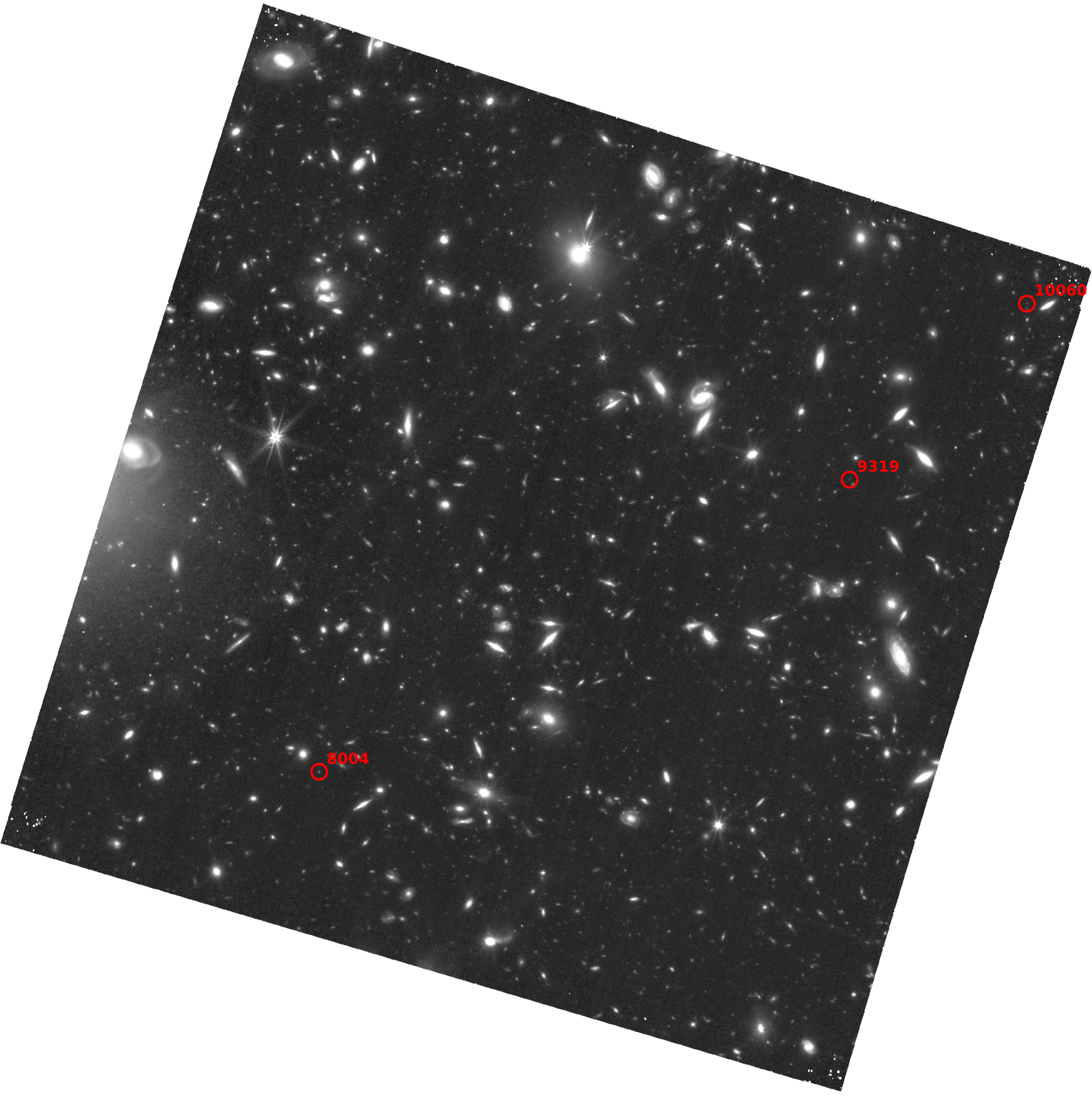}

\caption{
\JWST\ NIRCam detection image, comprised of a weighted sum of all NIRCam
LW images (F277W, F356W, F410M, F444W), of the nearby field (NIRCam A
module), centered $\sim2\farcm9$ from the \WHL\ cluster center. The
field of view is $\sim 2\farcm3 \times 2\farcm3$ and the image is
shown with north up and east left. The locations of the high-redshift
candidates are indicated with red circles. The location of the lensed
high-redshift candidate is shown in Figure~\ref{fig:nircam_cluster}.
\label{fig:nircam_par}}
\end{figure*}

\section{Observations} \label{sec:observations}

\begin{deluxetable*}{llcccrcccr}
\tablecaption{\HST\ and \JWST\ Exposure Times and Depths
\label{tbl:obs}}
\tablewidth{\columnwidth}
\tablehead{
\colhead{} &
\colhead{} &
\colhead{} &
\multicolumn{3}{c}{Cluster Field} &
\colhead{} &
\multicolumn{3}{c}{Parallel Field}
\\
\cline{4-6} \cline{8-10} \\
\colhead{} &
\colhead{} &
\colhead{Wavelength} &
\colhead{Exposure Time} &
\colhead{$m_{\mathrm{lim}}$\tnm{a}} &
\colhead{$f_{\mathrm{lim}}$\tnm{b}} &
\colhead{} &
\colhead{Exposure Time} &
\colhead{$m_{\mathrm{lim}}$\tnm{a}} &
\colhead{$f_{\mathrm{lim}}$\tnm{b}}
\\[-6pt]
\colhead{Camera} &
\colhead{Filter} &
\colhead{(\micron)} &
\colhead{(s)} &
\colhead{(nJy)} &
\colhead{(AB)} &
\colhead{} &
\colhead{(s)} &
\colhead{(nJy)} &
\colhead{(AB)}
}
\startdata
\HST\ ACS/WFC & F435W & 0.37--0.47 & ~\,2072   & 27.7 & ~\,29.5  & & \nodata & \nodata & \nodata \\
\HST\ ACS/WFC & F475W & 0.4--0.55  & ~\,3988   & 28.5 & ~\,14.9  & & \nodata & \nodata & \nodata \\
\HST\ ACS/WFC & F606W & 0.47--0.7  & ~\,2072   & 28.4 & ~\,16.5  & & \nodata & \nodata & \nodata \\
\HST\ ACS/WFC & F814W & 0.7--0.95  & 13326     & 28.8 & ~\,10.5  & & \nodata & \nodata & \nodata \\
\HST\ WFC3/IR & F105W & 0.9--1.2   & ~\,1411   & 27.9 & ~\,25.4  & & \nodata & \nodata & \nodata \\
\HST\ WFC3/IR & F110W & 0.9--1.4   & 10047     & 29.4 & ~\,6.17  & & \nodata & \nodata & \nodata \\
\HST\ WFC3/IR & F125W & 1.1--1.4   & ~\,~\,711 & 27.3 & 42.8    & & \nodata & \nodata & \nodata \\
\HST\ WFC3/IR & F140W & 1.2--1.6   & ~\,~\,711 & 27.5 & ~\,37.0  & & \nodata & \nodata & \nodata \\
\HST\ WFC3/IR & F160W & 1.4--1.7   & ~\,1961   & 27.9 & ~\,25.5  & & \nodata & \nodata & \nodata \\
\JWST\ NIRCam & F090W & 0.8--1.0   & ~\,4208   & 28.6 & ~\,13.3  & & ~\,2104 & 28.4    & ~\,16.0 \\
\JWST\ NIRCam & F115W & 1.0--1.3   & ~\,4208   & 28.6 & ~\,12.6  & & ~\,2104 & 28.4    & ~\,15.5 \\
\JWST\ NIRCam & F150W & 1.3--1.7   & ~\,2104   & 28.5 & ~\,14.1  & & ~\,2104 & 28.6    & ~\,13.2 \\
\JWST\ NIRCam & F200W & 1.7--2.2   & ~\,2104   & 28.7 & ~\,11.9  & & ~\,2104 & 28.7    & ~\,11.7 \\
\JWST\ NIRCam & F277W & 2.4--3.1   & ~\,4208   & 29.5 & ~\,5.6  & & ~\,2104 & 29.3    & ~\,6.7 \\
\JWST\ NIRCam & F356W & 3.1--4.0   & ~\,4208   & 29.7 & ~\,5.0  & & ~\,2104 & 29.4    & ~\,6.1 \\
\JWST\ NIRCam & F410M & 3.8--4.3   & ~\,2104   & 28.7 & ~\,11.6  & & ~\,2104 & 28.8    & ~\,11.2 \\
\JWST\ NIRCam & F444W & 3.8--5.0   & ~\,2104   & 29.1 & ~\,8.1  & & ~\,2104 & 29.1    & ~\,8.1
\enddata
\tablenotetext{a}{5$\sigma$ limiting AB magnitude in a $r=0\farcs1$ circular aperture}
\tablenotetext{b}{5$\sigma$ limiting flux in a $r=0\farcs1$ circular aperture}
\end{deluxetable*}

\subsection{\JWST\ Data}

We obtained \JWST\ NIRCam imaging of \WHL\ (GO 2282, PI Coe) in July
2022 and January 2023 as part of a program to further study Earendel
and the Sunrise Arc. The first epoch of NIRCam observations cover
eight filters (F090W, F115W, F150W, F200W, F277W, F356W, F410M, and
F444W) spanning $0.8 - 5.0~\micron$ with 2104 s of exposure time in
each filter. The NIRCam imaging was obtained over two $2\farcm26 \times
2\farcm26$ fields separated by 40.5\arcsec, covering 10.2 arcmin$^{2}$
in total. For the first epoch, the \WHL\ cluster was centered on NIRCam
module B while NIRCam module A obtained observations on a nearby field
centered $\sim$ 2\farcm9 northwest of the cluster center. For the
second epoch, the NIRCam observations cover four filters (F090W, F115W,
F277W, and F356W) with 2104 s of exposure time in each filter. The
observations for the second epoch were obtained 185\dg\ from the first
epoch. The cluster was again centered on NIRCam module B, while NIRCam
module A obtained observations on another nearby field southeast of the
cluster center. Because this second parallel field has imaging only in 4
filters, we do not use it in this analysis.

Each exposure uses the SHALLOW4 readout pattern with ten groups and
one integration. We use the INTRAMODULEBOX dither pattern with four
dithers to fill the 5\arcsec\ gaps in the short wavelength detectors
and to maximize the area with full exposure time. The dither pattern
also mitigates the effects of bad pixels and image artifacts and also
improves the spatial resolution of the resampled/drizzled images.

\subsection{\HST\ Data}

The RELICS \HST\ Treasury program \citep[GO 14096;][]{Coe2019} obtained
the first \HST\ imaging of the galaxy cluster \WHL\ in 2016 with three
orbits of ACS (F435W, F606W, and F814W) and two orbits of WFC3/IR
(F105W, F125W, F140W, and F160W) data spanning $0.4 - 1.7~\micron$.
Two follow-up \HST\ imaging programs (GO 15842 and GO 16668; PI: Coe)
have thus far obtained an additional 5 orbits of \HST\ ACS imaging in
F814W, 2 orbits in F475W, and 4 orbits with WFC3/IR in F110W. Two more
orbits of WFC3/IR F110W data are yet to be obtained from the Earendel
monitoring program (GO 16668). The \HST\ data cover only the cluster
field.




In total, the \JWST\ and \HST\ observations of \WHL\ include
imaging in 17 filters spanning $0.4 - 5.0~\micron$. We show color
images of the \JWST\ data in Figures~\ref{fig:nircam_cluster}
and \ref{fig:nircam_par}. The observations are summarized in
Table~\ref{tbl:obs}.

\section{Methods}

\subsection{Data Reduction}

We retrieved the pipeline-calibrated \HST\ data and the
\JWST\ level-2 imaging products and processed them using
\grizli\footnote{\url{https://github.com/gbrammer/grizli}} (version
1.8.12) reduction pipeline \citep{Grizli}. The calibrated \HST\ optical
and near-infrared data were obtained from Complete Hubble Archive for
Galaxy Evolution (CHArGE) \citep{Kokorev2022}. The \grizli\ image
reduction process is described in \cite{Valentino2023}. The \JWST\ data
were processed with version 1.9.6 of the calibration pipeline with
latest CRDS context \texttt{jwst\_1093.pmap}, which includes the most
recent photometric calibrations based on in-flight data.

For the \JWST\ data, the \grizli\ reduction pipeline applies
a correction to reduce the effect of $1/f$ noise and masks
``snowballs''\footnote{\url{https://jwst-docs.stsci.edu/data-artifacts-and-features/snowballs-artifact}}
that are are caused by large cosmic ray impacts to the
NIRCam detectors. The \grizli\ pipeline also includes a
correction for faint, diffuse stray light features, called
``wisps''\footnote{\url{https://jwst-docs.stsci.edu/jwst-near-infrared-camera/nircam-features-and-caveats/nircam-claws-and-wisps}}
that are present at the same detector locations in NIRCam images.
These stray-light features are most prominent in the NIRCam A3, B3,
and B4 detectors in the F150W and F200W data. A ``wisp'' template was
subtracted from each of these detectors for both the F150W and F200W
data.

The \grizli\ pipeline aligns the \HST\ and \JWST\ data to a common world
coordinate system registered to the GAIA DR3 catalogs \citep{Gaia_EDR3}.
The fully-calibrated images in each filter were combined and drizzled
to a common pixel grid using \astrodrizzle\ \citep{MultiDrizzle,
DrizzlePac}. The \HST\ and \JWST\ NIRCam long-wavelength (LW) filters
(F277W, F356W, F410M, and F444W) were drizzled to a grid of 0$\farcs$04
per pixel while the \JWST\ short-wavelength NIRCam filters (F090W,
F115W, F150W, and F200W) were drizzled to a grid of 0$\farcs$02 per
pixel.

These \grizli\ reduced images are available publicly, alongside images
and catalogs from other JWST programs with public
data.\footnote{\url{https://jwst-grizli.s3.amazonaws.com/sunrise-new/sunrise-v2_index.html}}

\subsection{Photometric Catalogs}

To produce the photometric catalogs, the NIRCam SW images were first
rebinned to a pixel scale of 0$\farcs$04 per pixel, placing the images
for all 17 filters on the same pixel grid. Sources were then identified
in a detection image comprised of a weighted sum of the F277W, F356W,
and F444W NIRCam LW images using \photutils\ \citep{Bradley2023}
image-segmentation tools. Visual inspection of the segmentation image
revealed a $2\farcs4$ long lensed arc that had been segmented into five
separate components. Therefore, we combined the separate arc segments
into a single source before performing photometry.

Source fluxes were measured with \photutils\ in flexible elliptical
Kron apertures with a scale factor of 1.5. The size of the elliptical
Kron aperture is calculated for each source by multiplying the Kron
scale factor by the Kron radius, which is calculated independently for
each source using the first-order moment of its flux distribution.
It has been shown that measuring colors in small elliptical
apertures accurately recovers the colors of distant galaxies
\citep{Finkelstein2022,Finkelstein2023}. We then performed a second run
of \photutils\ on the detection image using Kron apertures with a scale
factor of 2.5. An aperture correction to the total flux for the small
apertures was estimated as the ratio between the flux in the larger
aperture and that in the smaller aperture for each source. We then
applied this aperture correction the fluxes and uncertainties for all
filters.

\subsection{Photometric Redshifts} \label{sec:photoz}

We derive initial photometric redshifts using \eazypy\
\citep{Brammer2008}, which fits the observed photometry of each galaxy
using a set of templates added in a non-negative linear combination.
We use the photometry measured in elliptical Kron apertures with a
scale factor of 1.5. Both \JWST\ and \HST\ photometry is included in
the photometric redshift calculations for the \WHL\ cluster field,
while only \JWST\ photometry is used for the Module A field. The
photometric redshifts were calculated using a template set comprised
of the 12 ``tweak\_fsps\_QSF\_12\_v3'' templates derived from the
Flexible Stellar Population Synthesis (FSPS) library \citep{Conroy2009,
Conroy2010a, Conroy2010b}, which include a range of galaxy types (e.g.,
star-forming, quiescent, dusty) and realistic star formation histories
(e.g., bursty, slowly rising, slowly falling). To these FSPS templates,
we add six templates from \citep{Larson2022} that span bluer colors than
he fiducial FSPS templates. These additional templates were found to
provide better photometric-redshift accuracies for bluer galaxies at $z
> 9$ \citep{Larson2022}. We allow the redshifts to span from $0.1 < z <
20$, in steps of 0.01. Because we are just beginning to explore galaxies
at these epochs, the high-redshift luminosity function, especially at
the bright end, is not well-known at $z \ga 9$. Therefore, we adopt
a flat luminosity prior, similar to recent similar to recent \JWST\
high-redshift studies \citep[e.g.,][]{Finkelstein2022, Adams2023,
Finkelstein2023}, to prevent bias against the selection of bright
high-redshift galaxies.

\subsection{High-Redshift Candidate Selection} \label{sec:selection}

We select our initial sample of high-redshift candidate galaxies using
a combination of criteria using both signal-to-noise and photometric
redshift measurements. Measuring photometric redshifts using SED
fitting is a well-established method for selecting high-redshift
galaxy candidates that simultaneously uses the photometry in all bands
\citep[e.g.,][]{ Bradley2014, Salmon2020, Finkelstein2022, Naidu2022,
Adams2023, Donnan2023, Finkelstein2023}. The photometric signal-to-noise
(SNR) criteria are used to both ensure non-detections in filters
blueward of \Lya\ and to ensure robust photometric detections in
multiple filters redward of the \Lya\ break, which minimizes spurious
noise detections. We also visually inspect each candidate galaxy in each
filter image and its best-fit SED to remove detector artifacts and other
spurious sources such as diffraction spikes, misidentified parts of
larger galaxies, and spurious noise close to the detector edge.

We use a criteria similar to \cite{Finkelstein2023} to select
our initial sample of high-redshift candidates while minimizing
contamination from low-redshift interlopers:

\begin{itemize}
    \item A SNR of $<1.5$ to ensure non-detections blueward of \Lya\ in all of the following filters: F435W, F475W, F606W, F814W, and F090W.
    \item A SNR of $>5.5$ in at least two the following filters: F115W, F150W, F200W, F277W, F356W, and F444W to reduce spurious sources.
    \item Best-fit photometric redshift measured by \eazypy\ of $\zbest \ge 8.5$
    \item Integral of the \eazypy\ posterior redshift probability (\Pz) at $z > 8$ of $\int \Pzgtr{8}~\dz > 0.8$
    \item $\chi^2$ of the best-fit \eazypy\ spectral-energy distribution (SED) of $\chi^2 < 30$
\end{itemize}

We also require additional non-detections as a function of the redshift
selection window as follows. For the $9.7 \le z < 13$ sample, we require
the F115W SNR $< 1.5$. For the $z \ge 13$ sample, we require the F115W
and F150W SNR $< 1.5$. These filters are bluer than \Lya\ at these
corresponding redshifts.

Additionally, we ran \eazypy\ restricting the maximum redshift to $z
< 7$. We then calculate the difference of the best-fit $\chi^2$ for
these ``low-redshift" solutions and the best-fit $\chi^2$ for the
high-redshift solutions. For our high-redshift sample, we require a
conservative $\Delta \chi^2 > 9$, ruling out the low-redshift model at
$\ge 3\sigma$ significance \citep{Harikane2023}.

As a further check, we also calculated photometric redshifts using
\eazypy\ with the recently-added SFHZ templates. These templates have
redshift-dependent star formation histories (SFH) that disfavor star
formation starting earlier than the age of the universe at a given
epoch. We excluded candidates from our high-redshift sample where these
templates prefer a low-redshift ($\zbest < 8.5$) solution.

\section{Results and Discussion} \label{sec:results}

\subsection{High-Redshift Sample}

Our final sample consists of four high-redshift galaxy candidates. One
of the candidates lies in the \WHL\ cluster field, while the remaining
three are located in the nearby parallel field. As measured by \eazypy,
three of the candidates are at $8.5 \le z < 9.7$, while the remaining
candidate was identified in the $9.7 \le z < 13$ selection.

\begin{figure*}
\centering
\begin{center}
\centerline{\includegraphics[width=0.99\hsize]{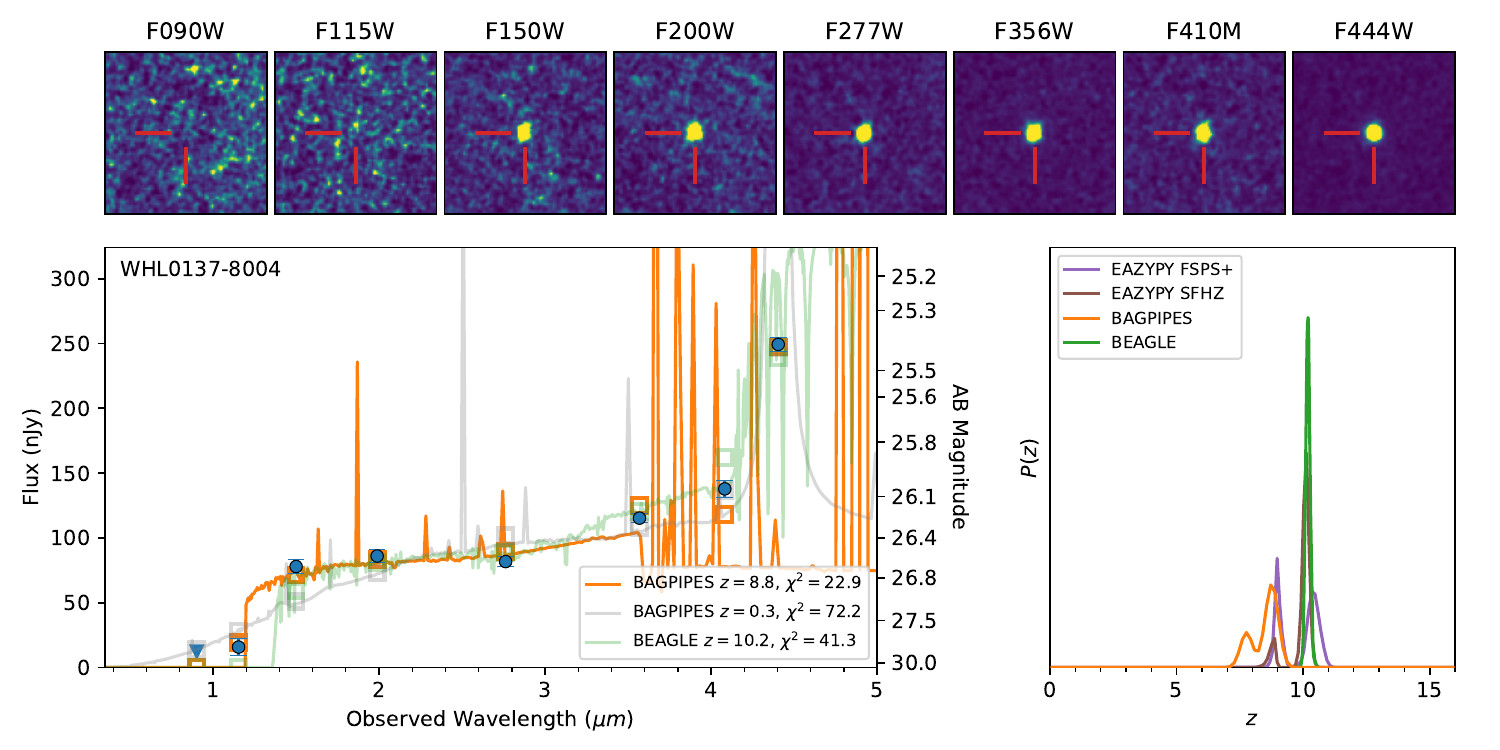}}
\end{center}

\begin{center}
\centerline{\includegraphics[width=0.99\hsize]{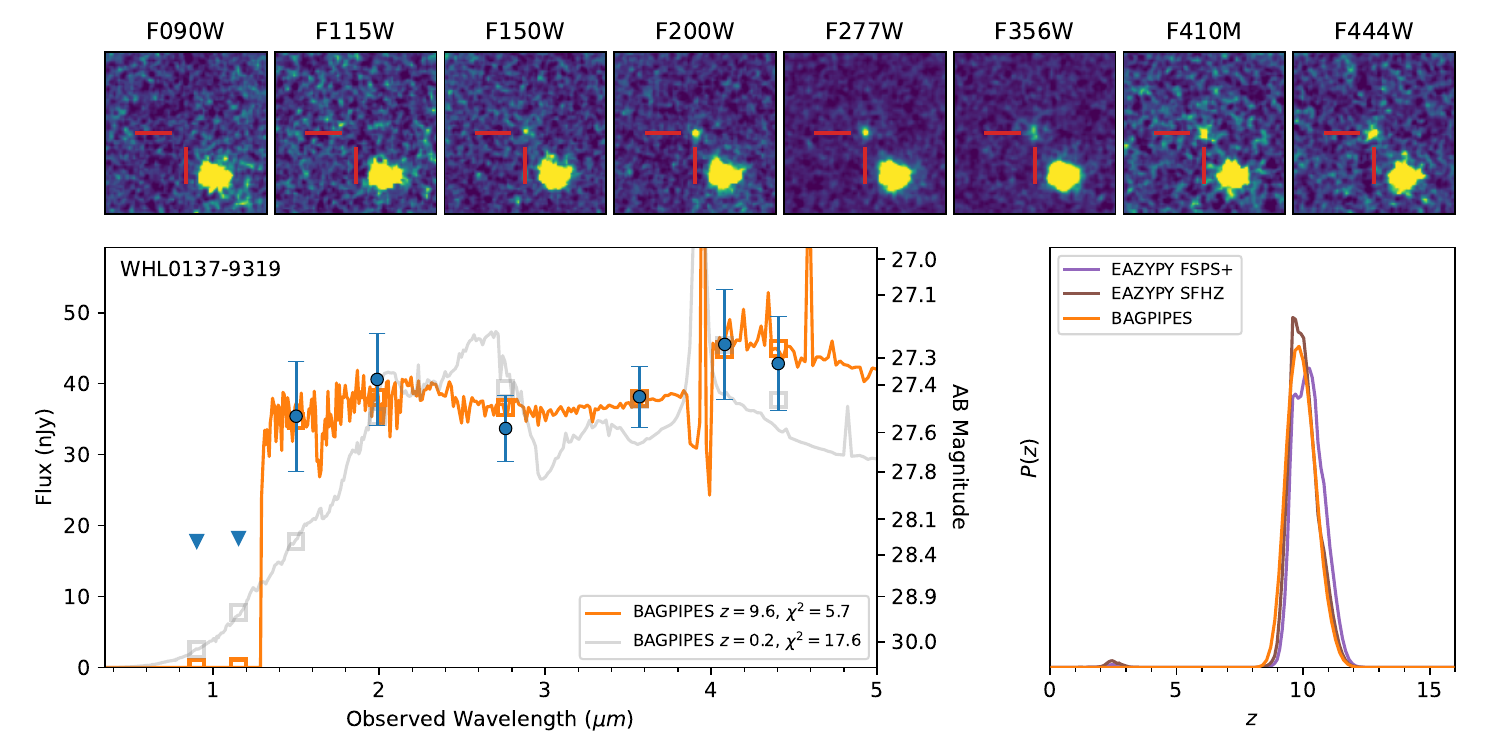}}
\end{center}

\caption{Cutout images, best-fit spectral energy distributions (SEDs), and posterior redshift distributions for the high-redshift galaxy candidates \idnum{08004} and \idnum{09319}.
{\bf Top panels:} $3\arcsec \times 3\arcsec$ \JWST\ cutout images spanning $0.9 - 4.5~\micron$ centered on each candidate.
{\bf Bottom-left panels:} Source photometry is shown as blue data points or triangle upper limits. Non-detections are plotted as upper limits at the $1\sigma$ level. The best-fit \bagpipes\ spectral energy distribution (SED) model at high-redshift ($z \ge 7$) is shown in orange with squares indicating the expected photometry in a given band. The best-fit \bagpipes\ SED for a low-redshift ($z < 7$) solution is shown in gray.
For \idnum{08004} we also show the best-fit \beagle\ SED model (see Section~\ref {sec:properties}) in green.
{\bf Bottom-right panels:}  Posterior probability distributions (\Pz) for the source photometric redshift derived using \eazypy\ (using both the FSPS+\cite{Larson2022} and SFHZ template sets), \bagpipes, and \beagle.}
\label{fig:candidates_a}
\end{figure*}

\begin{figure*}
\centering
\begin{center}
\centerline{\includegraphics[width=0.99\hsize]{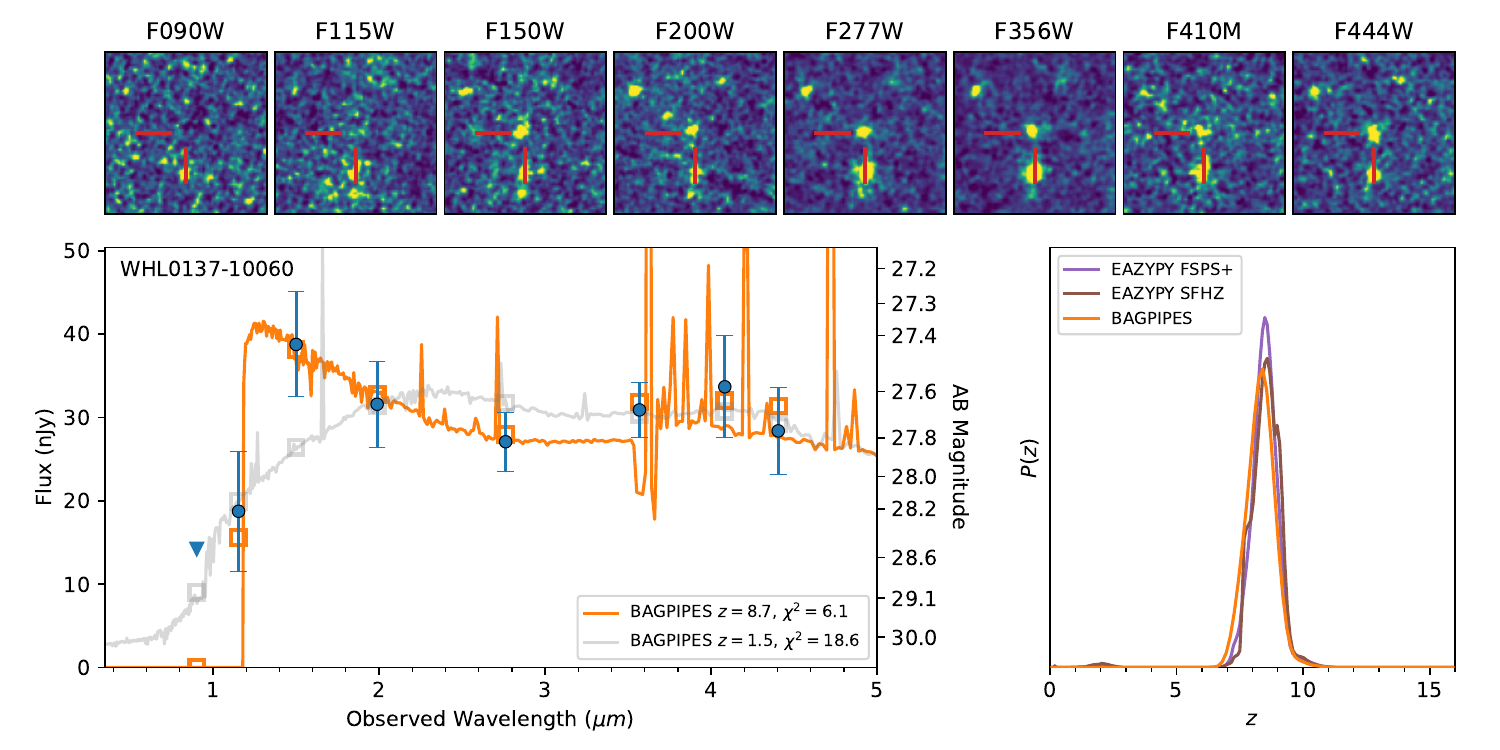}}
\end{center}

\begin{center}
\centerline{\includegraphics[width=0.99\hsize]{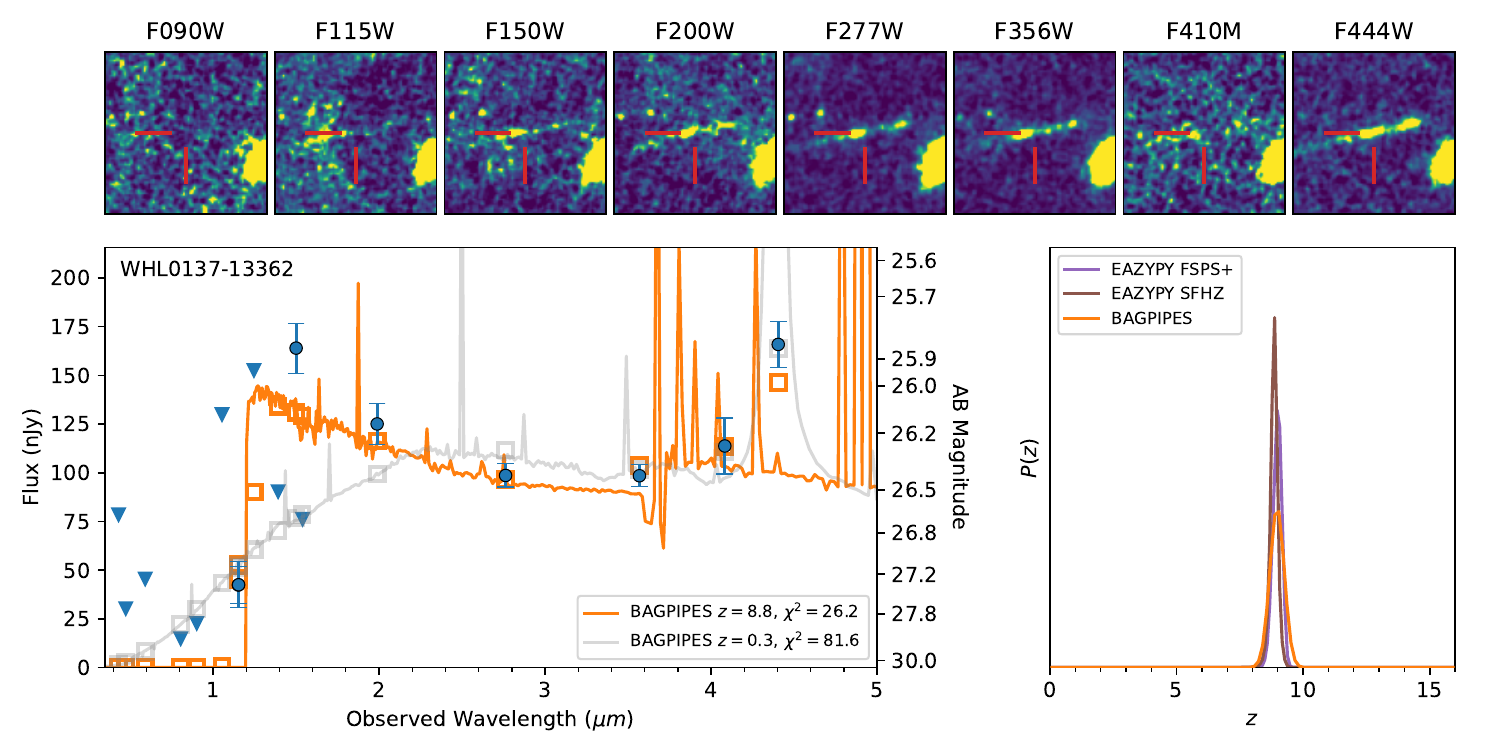}}
\end{center}

\caption{Same as Figure~\ref{fig:candidates_a}, but for
the high-redshift galaxy candidates \idnum{10060} and \idnum{13362}.}
\label{fig:candidates_b}
\end{figure*}

The measured (uncorrected for magnification) \JWST\ photometry of our
high-redshift candidates is presented in Table~\ref{tbl:jwstphot}.
For the lensed high-redshift candidate in the cluster field, we also
present its measured (uncorrected for magnification) \HST\ photometry
in Table~\ref{tbl:hstphot}. In Figures~\ref{fig:candidates_a} --
\ref{fig:candidates_b} we present $3\arcsec \times 3\arcsec$ cutout
images, the best-fit SEDs, and the posterior redshift distributions,
\Pz, for each candidate. The posterior redshift distributions
include plots for \eazypy\ (for both the FSPS+\citep{Larson2022}
and SFHZ template sets), \bagpipes, and \beagle. We plot in
Figures~\ref{fig:candidates_a} -- \ref{fig:candidates_b} the \bagpipes\
best-fit high-redshift ($z \ge 7$) SED along with the best-fit
low-redshift SED constrained with redshift $z < 7$.

We estimate the amplitude of the \Lya\ break in these candidates as 2.8,
$>1.5$, $>1.8$, and $>1.5$ magnitudes, which are significant breaks
(factors of $3.5 - 13$ in flux ratios). The upper limits were calculated
using the measured flux uncertainty for the non-detection blueward of
the break. Our measured break colors are larger than the break amplitude
of 0.5 mag used to select $9 < z < 11$ high-redshift candidates in
\citep{Atek2023}, comparable to the 1.7 mag used to select $9 < z <
11.5$ candidates in \citep{Castellano2022}, and consistent with the
1.5 ($z \sim 8$ selection) and 1.4 ($z \sim 10$ selection) break color
criterion in \citep{Bouwens2023}, who all used color-color selection
criteria to select their high-redshift candidates.

\tabletypesize{\tiny}
\begin{deluxetable*}{ccccccccccc}
\tablecolumns{11}
\tablecaption{\JWST\ photometry for the complete sample of high-redshift candidates
\label{tbl:jwstphot}}
\tablehead{\colhead{Object ID} & \colhead{RA} & \colhead{Dec} & \colhead{F090W} & \colhead{F115W} & \colhead{F150W} & \colhead{F200W} & \colhead{F277W} & \colhead{F356W} & \colhead{F410M} & \colhead{F444W}\\ \colhead{ } & \colhead{ } & \colhead{ } & \colhead{$\mathrm{nJy}$} & \colhead{$\mathrm{nJy}$} & \colhead{$\mathrm{nJy}$} & \colhead{$\mathrm{nJy}$} & \colhead{$\mathrm{nJy}$} & \colhead{$\mathrm{nJy}$} & \colhead{$\mathrm{nJy}$} & \colhead{$\mathrm{nJy}$}}
\startdata
WHL0137-08004 & $24.34989786$ & $-8.41981970$ & $-1.1 \pm 6.1$ & $15.8 \pm 6.4$ & $77.9 \pm 5.7$ & $86.0 \pm 4.8$ & $81.7 \pm 3.7$ & $115.3 \pm 3.6$ & $137.8 \pm 6.4$ & $249.3 \pm 5.5$ \\
WHL0137-09319 & $24.32694954$ & $-8.40731680$ & $-4.0 \pm 8.9$ & $-19.9 \pm 9.1$ & $35.4 \pm 7.7$ & $40.6 \pm 6.5$ & $33.7 \pm 4.6$ & $38.2 \pm 4.3$ & $45.5 \pm 7.7$ & $42.8 \pm 6.6$ \\
WHL0137-10060 & $24.31928158$ & $-8.39978978$ & $-16.3 \pm 7.1$ & $18.7 \pm 7.2$ & $38.8 \pm 6.3$ & $31.6 \pm 5.2$ & $27.1 \pm 3.6$ & $30.9 \pm 3.3$ & $33.7 \pm 6.1$ & $28.4 \pm 5.2$ \\
WHL0137-13362\tnm{a} & $24.35501927$ & $-8.44791625$ & $-16.0 \pm 11.3$ & $42.5 \pm 11.8$ & $164.0 \pm 12.9$ & $125.1 \pm 10.7$ & $98.6 \pm 5.9$ & $98.5 \pm 5.6$ & $113.7 \pm 14.4$ & $165.9 \pm 11.6$ \\
\enddata
\tablecomments{Observed fluxes, uncorrected for magnification. $\mAB = 31.4 - 2.5 \log(f_{\nu} / {\rm nJy}).$}
\tablenotetext{a}{Lensed candidate with a magnification of $\mu = 8^{+12}_{-6}$. Other candidates (located in the nearby field) are estimated to have magnifications of $\mu \le 1.1$.}
\end{deluxetable*}

\begin{deluxetable*}{ccccccccccc}
\tablecolumns{11}
\tablecaption{\HST\ Photometry of the lensed high-redshift candidate
\label{tbl:hstphot}}
\tablehead{\colhead{Object ID} & \colhead{F105W} & \colhead{F110W} & \colhead{F125W} & \colhead{F140W} & \colhead{F160W} & \colhead{F435W} & \colhead{F475W} & \colhead{F606W} & \colhead{F814W} & \colhead{$\mu$}\\ \colhead{ } & \colhead{$\mathrm{nJy}$} & \colhead{$\mathrm{nJy}$} & \colhead{$\mathrm{nJy}$} & \colhead{$\mathrm{nJy}$} & \colhead{$\mathrm{nJy}$} & \colhead{$\mathrm{nJy}$} & \colhead{$\mathrm{nJy}$} & \colhead{$\mathrm{nJy}$} & \colhead{$\mathrm{nJy}$} & \colhead{ }}
\startdata
WHL0137-13362 & $83.7 \pm 46.3$ & $42.4 \pm 9.6$ & $70.7 \pm 81.8$ & $21.5 \pm 68.8$ & $28.5 \pm 47.7$ & $-36.7 \pm 39.2$ & $12.5 \pm 17.7$ & $-5.3 \pm 22.8$ & $1.7 \pm 13.1$ & $8^{+12}_{-6}$
\enddata
\tablecomments{Observed fluxes, uncorrected for magnification. $\mAB = 31.4 - 2.5 \log(f_{\nu} / {\rm nJy}).$}
\end{deluxetable*}

\tabletypesize{\normalsize}

\subsection{Magnifications}
\label{sec:magnifications}

To estimate source magnifications, we use the lens models
constructed to analyze Earendel and the Sunrise Arc published
in \cite{Welch2022_earendel} and which were made publicly
available.\footnote{\url{https://relics.stsci.edu/lens_models/outgoing/whl0137-08/}}
These models were generated using four independent lens modeling
software packages: Light-Traces-Mass \citep[LTM,][]{Zitrin09, Zitrin15,
Broadhurst05}, Glafic \citep{Oguri2010}, WSLAP+ \citep{Diego05wslap,
Diego07wslap2}, and Lenstool \citep{JulloLenstool07, JulloLenstool09}.
Due to a lack of multiply-imaged sources in this cluster, the slope
of the lensing potential in these models varies by a factor of six,
which adds considerable uncertainty to our magnification estimates. For
further details about each model, please see \cite{Welch2022_earendel}.
The ratio of the source-plane area at $z = 9$ of the lensed cluster
field to the nearby parallel field ranges from 0.24 to 0.85 (ratios of
0.24, 0.29, 0.68, and 0.85 for the four models). The large range is
reflective of the uncertainty in the slope of the lensing potential.

One candidate in our sample is strongly lensed by the \WHL\ galaxy
cluster, while the other three candidates, located in the nearby NIRCam
module, are expected to have only weak magnifications of $\mu \le 1.1$.
The lensed candidate is \idnum{13362} with a \zphot $\sim 9$. At this
redshift, our lens models yield a magnification in the range from $\mu
= 2.4 - 20.2$, with an arithmetic mean value of $\mu = 7.9$. The mean
magnification for this candidate is quoted in Tables~\ref{tbl:hstphot},
~\ref{tbl:bagpipes_properties}, and \ref{tbl:beagle_properties}.

\subsection{Galaxy Sizes}
\label{sec:sizes}

We measure the half-light radius of the three unlensed candidates in the
sample from the deeper detection image using the SourceCatalog class
in \photutils. The half-light radius $r_{h}$ is calculated relative to
the ``total" flux measured in elliptical Kron apertures with a Kron
factor of 2.5. The derived sizes for \idnum{8004}, \idnum{9319}, and
\idnum{10060} are 3.3, 4.0, and 3.3 pixels (0.13\arcsec, 0.16\arcsec,
and 0.13\arcsec), respectively. Therefore, all three sources are
spatially resolved, being larger than the NIRCam F444W PSF full-width
at half maximum of 0\farcs145 (0\farcs0725 radius). The lensed sourced,
\arcid, is also spatially resolved, stretched into a 2\farcs4-long arc
by the effects of gravitational lensing. The morphology of the arc is
discussed in more detail in Section~\ref{sec:z9arc}.

Assuming the estimated \bagpipes\ photometric redshifts (see
Table~\ref{tbl:bagpipes_properties}) of the three unlensed sources,
they have physical sizes of 0.60, 0.67, and 0.61~kpc, respectively.
These sizes are comparable to the galaxy sizes found in the GLASS-JWST
survey ranging from $0.17 - 2.01$~kpc in F444W \citep{Yang2022} and
the CEERS survey, with sizes ranging from $0.05 - 1.1$~kpc in F200W
\citep{Finkelstein2023}.

\subsection{Spectral Energy Distribution (SED) Fitting} \label{sec:sedfitting}

\subsubsection{\bagpipes}

For each galaxy in our high-redshift sample, we estimate its physical
properties using SED fitting. We performed SED fitting using the
Bayesian Analysis of Galaxies for Physical Inference and Parameter
EStimation \citep[\bagpipes;][]{bagpipes} Python package and the
BayEsian Analysis of GaLaxy sEds \citep[\beagle;][]{beagle} tool with
redshift as a free parameter.

\bagpipes\ generates model galaxy spectra over the multidimensional
space of physical parameters and fits these to the photometric data
using the \multinest\ nested sampling algorithm \citep{Feroz2008,
Feroz2009, Feroz2013}. \bagpipes\ uses the stellar population synthesis
models from the 2016 version of the BC03 \citep{BC03} models. These
models were generated using a \cite{Kroupa2002} initial mass function
(IMF) and include nebular line and continuum emission based on \cloudy\
\citep{Ferland2013}, with the logarithm of ionization parameter ($\log$
U) allowed to vary between $-4$ to $-2$. We perform our SED fitting
using a delayed exponentially declining SFH where the star formation
rate (SFR) is of the form SFR$(t) \propto t \ \exp{(-t/\tau)}$. Models
assuming a constant star formation rate yield younger ages and higher
sSFRs, as discussed below in \S\ref{sec:properties}.

For SED fits constrained to be at low redshift ($z < 7$, we assume
a Calzetti law \citep{Calzetti2000} for dust attenuation. For SED
fits constrained to be a high redshift ($z > 7$), we assume a Small
Magellanic Cloud (SMC) dust law \citep{Salim2018}. For both cases,
we also include a second component to the dust model that includes
birth-cloud dust attenuation that is a factor of two larger around \HII\
regions as in the general ISM within the galaxy's first 10 Myr. We
allow dust extinction to range from $A_V$ = $0 - 5$ magnitudes and we
vary metallicity in logarithmic space from $\log Z/\Zsun = 0.005 - 5$.
Formation ages vary from 1 Myr to the age of the universe.

\subsubsection{\beagle}

We also perform SED fitting on each candidate galaxy using the \beagle\
tool \citep{beagle} with simplified version of the configuration,
fit parameters, and parameter space used in \citet{Atek2023} and
\citet{Furtak2023}. \beagle\ uses SED templates by \cite{Gutkin2016},
which also combine the 2016 version of the BC03 stellar population
synthesis models with \cloudy\ to account for nebular emission. The
templates include ionization parameters varying from $-4$ to $-1$.
These templates all assume a \cite{Chabrier2003} IMF and model the
intergalactic attenuation using the \cite{Inoue2014} attenuation
curves. As with \bagpipes, we assume a delayed exponential SFH, but
with the possibility of an ongoing star-burst over the last 10\,Myr.
This allows for maximum flexibility of the SFH to be either rising or
declining with a maximum at $t = \tau$. We account for dust attenuation
by assuming an SMC-like dust attenuation law \citep{Pei1992}, which
has been found to fit high-redshift galaxies best at low metallicities
\citep{Capak2015, Reddy2015, Reddy2018, Shivaei2020}. Due to the
relatively large number of free parameters, we fix the metallicity
to $Z=0.1\,\mathrm{Z}_{\odot}$ while the stellar mass, current
SFR, maximal stellar age and dust attenuation are allowed to vary
freely in the ranges $\log(M_{\star}/\mathrm{M}_{\odot})\in[6,
11]$, $\log(\psi/\mathrm{M}_{\odot}\,\mathrm{yr}^{-1})\in[-4, 4]$,
$\log(t_{\mathrm{age}}/\mathrm{yr})\in[6, t_{\mathrm{Universe}}]$ and
$A_V\in[0, 3]$ respectively.

\tabletypesize{\small}
\begin{deluxetable*}{cccccccccc}
\tablecolumns{10}
\tablecaption{\bagpipes\ photometric redshifts and physical properties of the high-redshift galaxy candidates
\label{tbl:bagpipes_properties}}
\tablehead{
\colhead{Object ID} &
\colhead{\mumean\tnm{a}} &
\colhead{\zphot\tnm{b}} &
\colhead{\zphot\tnm{c}} &
\colhead{\logmstar} &
\colhead{SFR\tnm{d}} &
\colhead{$\log$ sSFR/Gyr$^{-1}$} &
\colhead{Age\tnm{e}} &
\colhead{$A_V$} &
\colhead{\tform\tnm{f}}
\\[-0.5em]
\colhead{} &
\colhead{} &
\colhead{} &
\colhead{$z > 7$} &
\colhead{} &
\colhead{$\Msun/yr$} &
\colhead{} &
\colhead{Myr} &
\colhead{mag} &
\colhead{Myr}
}
\startdata
WHL0137-08004 & \nodata & $8.7_{-0.3}^{+0.1}$ & $8.8_{-1.0}^{+0.1}$ & $8.39_{-0.07}^{+0.04}$ & $2.5_{-0.4}^{+0.3}$ & $1.0_{-0.0}^{+0.0}$ & $2_{-1}^{+1}$ & $0.15_{-0.02}^{+0.02}$ & 556 \\
WHL0137-09319 & \nodata & $9.9_{-0.8}^{+1.1}$ & $9.9_{-0.8}^{+1.2}$ & $8.74_{-0.27}^{+0.18}$ & $4.5_{-1.3}^{+1.4}$ & $0.9_{-0.2}^{+0.1}$ & $79_{-49}^{+51}$ & $0.09_{-0.05}^{+0.06}$ & 392 \\
WHL0137-10060 & \nodata & $8.4_{-1.2}^{+1.0}$ & $8.3_{-1.0}^{+0.9}$ & $8.43_{-0.24}^{+0.19}$ & $2.3_{-0.5}^{+0.4}$ & $1.0_{-0.2}^{+0.1}$ & $73_{-42}^{+67}$ & $0.04_{-0.03}^{+0.04}$ & 518 \\
WHL0137-13362 & $8_{-6}^{+12}$ & $9.0_{-0.3}^{+0.2}$ & $9.0_{-0.3}^{+0.3}$ & $8.31_{-0.17}^{+0.08}$ & $1.2_{-0.1}^{+0.1}$ & $0.8_{-0.1}^{+0.2}$ & $132_{-56}^{+37}$ & $0.02_{-0.01}^{+0.02}$ & 406 \\
\enddata
\tablecomments{Physical parameter results are quoted for high-redshift solutions restricting $z > 7$. We quote the median and the $1\sigma$ range of the joint posterior distributions for each galaxy. We have modeled star formation histories as exponential delayed $\tau$ model. If constant star formation histories are assumed, age estimates decrease and sSFR increases. For the lensed source, stellar masses and SFRs are corrected for the mean magnification. Multiply these values by $\mu_{\rm mean} / \mu$ to apply a different magnification. We did not propagate magnification uncertainties to those parameter uncertainties.}
\tablenotetext{a}{Mean magnification and uncertainties based on multiple independent lens models. Candidates in the nearby field are estimated to have magnifications of $\mu \le 1.1$.}
\tablenotetext{b}{Photometric redshift with 2$\sigma$ uncertainties, using the Calzetti dust law \citep{Calzetti2000}.}
\tablenotetext{c}{Photometric redshift restricted to $z > 7$ with 2$\sigma$ uncertainties, using the SMC dust law \citep{Salim2018}.}
\tablenotetext{d}{Star formation rate during the past 100 Myr.}
\tablenotetext{e}{Mass-weighted age for the delayed $\tau$ star formation history.}
\tablenotetext{f}{Formation time in Myr after the Big Bang based on the mass-weighted age.}
\end{deluxetable*}
\tabletypesize{\normalsize}

\tabletypesize{\small}
\begin{deluxetable*}{cccccccccc}
\tablecolumns{10}
\tablecaption{\beagle\ photometric redshifts and physical properties of the high-redshift galaxy candidates
\label{tbl:beagle_properties}}
\tablehead{
\colhead{Object ID} &
\colhead{\mumean\tnm{a}} &
\colhead{\zphot} &
\colhead{\logmstar} &
\colhead{SFR} &
\colhead{$\log$ sSFR/Gyr} &
\colhead{Age\tnm{b}} &
\colhead{$A_V$} &
\colhead{$\beta$\tnm{c}} &
\colhead{$M_{UV}$\tnm{d}}
\\[-0.5em]
\colhead{} &
\colhead{} &
\colhead{} &
\colhead{} &
\colhead{$\Msun$/yr} &
\colhead{} &
\colhead{Myr} &
\colhead{mag} &
\colhead{} &
\colhead{}
}
\startdata
WHL0137-08004 & \nodata & $10.2_{-0.1}^{+0.1}$ & $10.19_{-0.04}^{+0.05}$ & $0.1_{-0.1}^{+0.4}$ & $-2.2_{-0.7}^{+0.7}$ & $236_{-46}^{+44}$ & $0.02_{-0.01}^{+0.02}$ & $-1.5 \pm 0.1$ & $-22.4$ \\
WHL0137-09319 & \nodata & $10.1_{-0.5}^{+0.5}$ & $8.66_{-0.50}^{+0.31}$ & $5.4_{-5.1}^{+2.5}$ & $1.0_{-1.5}^{+0.7}$ & $33_{-19}^{+73}$ & $0.11_{-0.06}^{+0.09}$ & $-2.0 \pm 0.2$ & $-19.8$ \\
WHL0137-10060 & \nodata & $8.3_{-0.6}^{+0.6}$ & $8.37_{-0.17}^{+0.10}$ & $0.3_{-0.3}^{+1.4}$ & $0.1_{-1.0}^{+0.9}$ & $17_{-4}^{+10}$ & $0.09_{-0.06}^{+0.08}$ & $-2.2 \pm 0.2$ & $-19.0$ \\
WHL0137-13362 & $8_{-6}^{+12}$ & $9.0_{-0.2}^{+0.2}$ & $8.30_{-0.13}^{+0.12}$ & $0.4_{-0.4}^{+3.0}$ & $-0.6_{-1.2}^{+1.1}$ & $34_{-10}^{+15}$ & $0.02_{-0.01}^{+0.03}$ & $-2.6 \pm 0.1$ & $-18.7$ \\
\enddata
\tablecomments{Results are quoted as the median and the $1\sigma$ range of the joint posterior distributions for each galaxy. For the lensed source, stellar masses and SFRs are corrected for magnification. Multiply these values by $\mu_{\rm mean} / \mu$ to apply a different magnification. We did not propagate magnification uncertainties to those parameter uncertainties.}
\tablenotetext{a}{Mean magnification and uncertainties based on multiple independent lens models. Candidates in the nearby field are estimated to have magnifications of $\mu \le 1.1$.}
\tablenotetext{b}{Mass-weighted age in Myr.}
\tablenotetext{c}{Rest-frame UV slope.}
\tablenotetext{d}{Rest-frame absolute UV magnitude in the band that contains 1500~\AA\ at the galaxy's photometric redshift.}
\end{deluxetable*}

\tabletypesize{\normalsize}


\subsection{Physical Properties} \label{sec:properties}

The derived physical properties for our candidate high-redshift
galaxies using \bagpipes\ and \beagle\ are presented in
Tables~\ref{tbl:bagpipes_properties} and \ref{tbl:beagle_properties},
respectively. The \bagpipes\ physical parameter results are quoted for
high-redshift solutions restricted to $z > 7$. For \idnum{13362}, we
divide by its mean magnification (see Section~\ref{sec:magnifications})
to calculate intrinsic stellar mass and SFR.

We estimate intrinsic stellar masses of \logmstar\ $\sim$ $8.3 - 8.7$
for all galaxies with \bagpipes\ and all but one with \beagle. The
\bagpipes\ SFRs range from $\sim 1.2 - 4.5$~\sfr, with specific star
formation rates (sSFRs) of $\sim 1$ Gyr$^{-1}$ ($0.8 - 1.0$ Gyr$^{-1}$).
The \beagle\ SFRs range from $\sim 0.1 - 5.4$~\sfr, with sSFRs of $0.01
- 10$ Gyr$^{-1}$. Note that \bagpipes\ considers the most recent 100 Myr
of star formation while \beagle\ considers the most recent 10 Myr (sSFR
results capped at 100 Gyr$^{-1}$).

In most cases, the SED fitting reveals relatively young ages of $<132$
Myr. The exception is the \beagle\ fit for \idnum{08004} (discussed
below), which has a age of 236 Myr. Median age estimates from \bagpipes\
are typically $\sim$76 Myr, while \beagle\ median ages are typically
younger $\sim$34 Myr. Switching \bagpipes\ to a constant star formation
history (CSFH) also results in younger median ages typically $\sim$22
Myr.

For all candidates, we also find low dust content with $A_V < 0.15$, as
expected due to the relatively blue rest-frame UV slopes in our sample
of $\beta = -1.5$ to $-2.6$ (see Table~\ref{tbl:beagle_properties}).

The \beagle\ SED fits of \idnum{08004} have the largest stellar mass
with \logmstar = $10.19_{-0.04}^{+0.05}$ coupled with the lowest star
formation rate of $0.1_{-0.1}^{+0.4}$ \sfr. This is a result of \beagle\
fitting the red F410M $-$ F444W = 0.6 color as a Balmer break (see
Figure~\ref{fig:candidates_a}; top), with an older mass-weighted age of
$236_{-46}^{+44}$ Myr. On the other hand, \bagpipes\ fits this galaxy as
an extremely young ($2_{-1}^{+1}$ Myr) galaxy with a high sSFR rate of
$\sim 10$ Gyr$^{-1}$ and strong inferred \OIII+\Hb\ emission (rest-frame
equivalent width of $\sim 2000$~\AA). The \bagpipes\ fit yields a more
typical mass of \logmstar = $8.39_{-0.07}^{+0.04}$.

The lensed galaxy \idnum{13362} also has a red F410M $-$ F444W color of
0.4. For this candidate, both \bagpipes\ and \beagle\ fit this galaxy
with a SED template containing strong \OIII+\Hb\ optical emission lines.
\beagle\ gives a young age of 34 Myr, while \bagpipes\ yields a slightly
older age of 132 Myr.

\subsubsection{SED Fitting Limitations}

It is important to note that these physical property results rely on
SED fitting to the broadband photometry, which is primarily in the
rest-frame UV of these candidate high-redshift galaxies. This rest-frame
wavelength regime is not ideally suited to investigate galaxy parameters
because it primarily probes very massive and short-lived stars in a
galaxy, which may not compose the bulk of its stellar mass. Because
of limited rest-frame optical photometry, we expect to SED fitting to
underestimate the stellar masses. In particular, \cite{Furtak2021}
showed that SED-fitting to only UV photometry can underestimate stellar
masses by up to 0.6 dex. Likewise, fitting primarily to rest-frame UV
photometry can lead to some degeneracies between the stellar mass, SFR,
and age \citep{Furtak2023}.

While the SED fitting results of \bagpipes\ and \beagle\ are relatively
consistent, they differ significant for \idnum{8004}, leading to
completely different interpretations. As discussed above, \bagpipes\
fits this galaxy with very strong \OIII+\Hb\ optical emission lines,
while \beagle\ fits this galaxy with a strong Balmer break. These
fitting differences result in highly uncertain physical properties
for this galaxy, with the stellar mass differing by 1.4 dex, the SFR
differing by 2.4 \sfr, and the mass-weighted age differing by 234 Myr
(either extremely young or relatively old).

We also emphasize that given the large magnification uncertainties
for \arcid, due to the lack of a multiply-imaged sources in the
\WHL\ cluster, the intrinsic stellar mass and SFR of this galaxy
are highly uncertain. As a result, in addition to the SED fitting
uncertainties shown in Tables~\ref{tbl:bagpipes_properties} and
\ref{tbl:beagle_properties}, the derived stellar mass and SFR can vary
systematically by factors of $0.75 - 1.5$.

\begin{figure}
\includegraphics[width=0.99\hsize]{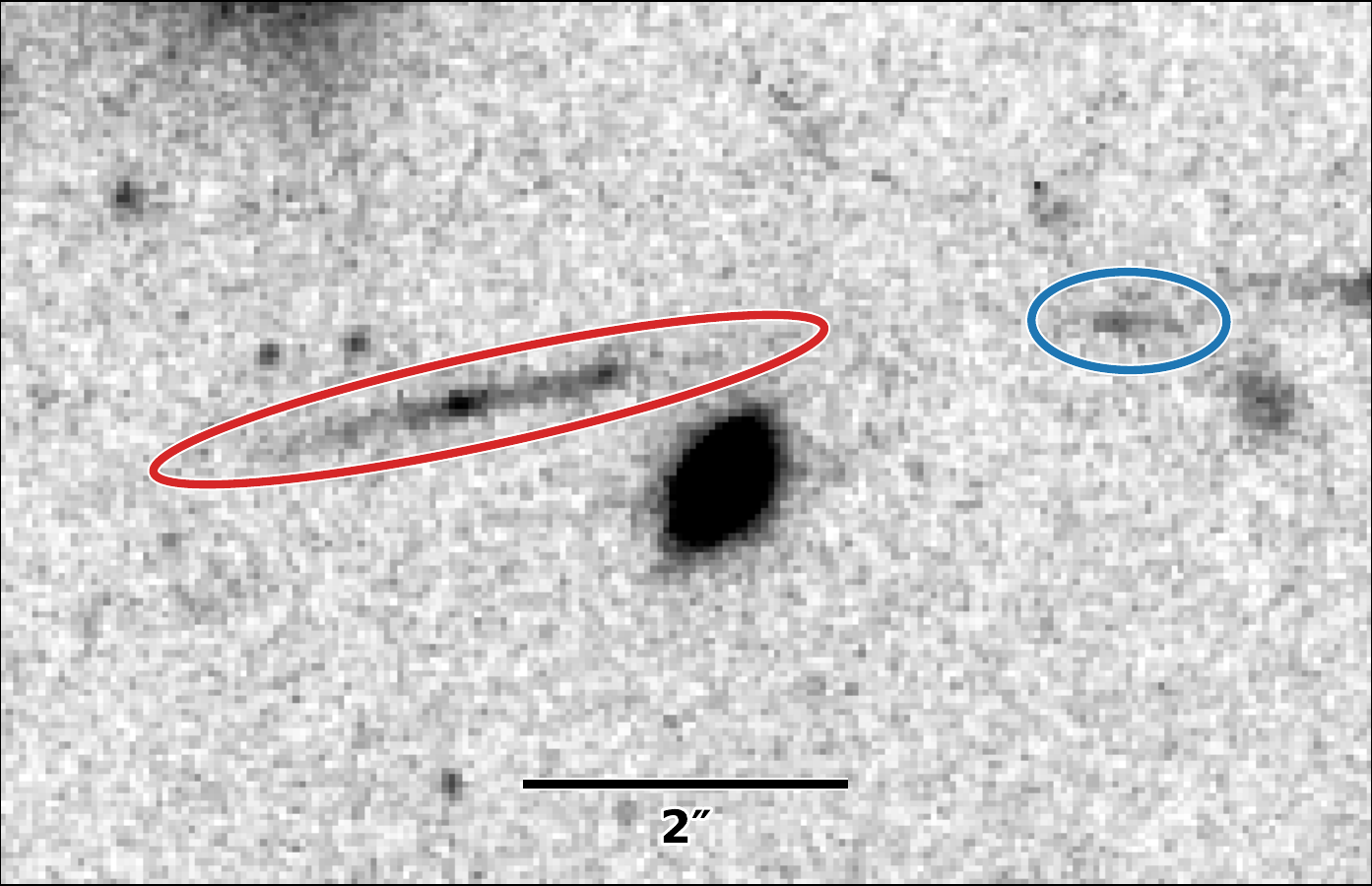}
\caption{$8\farcs5 \times 5\farcs5$ cutout image from the \JWST\
NIRCam detection image showing the $z \sim 9$ arc \arcid\ (red outline),
with a magnification of $\mu \sim 8$. The arc is 2\farcs4 long, has at
least two bright star-forming knots, and is the most-spatially resolved
arc at $z \sim 9$ known to date. A potential fainter counterimage of
the arc, with a similar color and photometric redshift, is also shown
outlined in blue.}
\label{fig:z9arc}
\end{figure}

\subsection{A gravitationally-lensed arc at \texorpdfstring{$z \sim 9$}{z ~ 9}}
\label{sec:z9arc}

The lensed high-redshift candidate, \arcid, is stretched into an
arc 2\farcs4 long by the effects of strong gravitational lensing.
The arc has at least two bright knots of unevenly distributed star
formation (see Figure~\ref{fig:z9arc}). This candidate has a lensed
F200W AB magnitude of $26.2 \pm 0.1$. Assuming a magnification of $\mu
= 7.9$ ($\mu = 2.4 - 20.2$; see Section~\ref{sec:magnifications}), its
intrinsic F200W AB magnitude is $28.4 \pm 0.1$.

The \bagpipes\ and \beagle\ SED fitting results for \arcid\ yield a
photometric redshift of $z = 9.0_{-0.3}^{+0.2}$. The redshift posterior
distribution $P(z)$ (see Figure~\ref{fig:candidates_b}) for all SED
fitting codes, including \eazypy, shows a very narrow distribution
peaked at $z\sim9$.

Additional NIRSpec multi-object spectroscopy observations using the
microshutter assembly were obtained in December 2022 (GO 2282: PI Coe).
While these observations were obtained primarily to study Earendel and
the Sunrise Arc in more detail, we were also able to put a slitlet on
the $z \sim 9$ arc \arcid. The NIRSpec prism spectroscopy covers $0.6
- 5.3~\micron$ with $R \sim 30-300$ with 3720~s of exposure time. The
NIRSpec data reveal \Hb\ and \OIIIww\ emission lines at $z = 8.22$,
confirming this high-redshift galaxy (Vikaeus et al. in prep).

After correcting for magnification, the \bagpipes\ results yield a
stellar mass of \logmstar = $8.31_{-0.17}^{+0.08}$ and a star formation
rate of $1.2_{-0.1}^{+0.1}$ \sfr. The galaxy is relatively young with
a mass-weighted age of $132_{-56}^{+37}$ Myr and a formation redshift
of $\zform = 11.1$ ($\tform = 406$~Myr). The \beagle\ results yield a
nearly identical stellar mass of \logmstar = $8.30_{-0.13}^{+0.12}$
and a lower star formation rate of $0.4_{-0.4}^{+3.0}$ \sfr. \beagle\
gives an even younger mass-weighted age of $34_{-10}^{+15}$ Myr and a
formation redshift of $\zform = 9.4$ ($\tform = 504$~Myr).

With a photometric redshift of $\zphot \sim 9$ ($\zspec = 8.22$),
\arcid\ is the most spatially-resolved galaxy at this redshift known to
date. Only SPT0615-JD1, the gravitationally lensed 2\farcs5 long arc,
is more distant at $z \sim 10$ \citep{Salmon2018}. We show a $8\farcs5
\times 5\farcs5$ cutout image of \arcid\ from the \JWST\ detection image
in Figure~\ref{fig:z9arc}. Assuming a primarily linear and tangential
magnification of $\mu \sim 8$, the intrinsic size of the arc is 1.4~kpc
at $z = 8.22$. Given the F150W point-spread function full width at half
maximum of $\sim$ 0.05\arcsec\ ($\sim 237$ pc at $z = 8.22$), we can
resolve $\sim$30 pc scales in this $z = 8.22$ galaxy.

Unfortunately, \WHL\ has no multiple image constraints in the northeast
section of the cluster, where we find this $z \sim 9$ arc. This adds
considerable uncertainty to predictions of counter image locations.
While some models (Lenstool) predict two merging images of the arc,
and a third image near the cluster center, other models (LTM) predict
no counter images. Although it is not the case with the current lens
models, it is possible that the two knots in the arc could be multiple
images if the critical curve happens to pass through the arc. Despite
the lensing uncertainties, we have identified a promising counter image
(\idnum{06156}) $4\arcsec$ to the west of the arc (RA = 24.3539132\dg,
Dec = $-$8.44778165\dg, J2000) with a similar color. While this source
is 1.8 mag fainter than the $z \sim 9$ arc, \eazypy estimates its
$\zphot = 8.2_{-0.4}^{+0.6}$, which very closely agrees with the
redshift $\zspec = 8.22$ of the main arc. Additional lens modeling is
ongoing to further investigate this possibility.

\subsection{Number Counts}

In this dataset, the lensed field yielded one candidate, fewer than the
three identified in the nearby blank field. While this difference may
simply be due to small number statistics, we explore other possible
reasons.

As discussed in \S\ref{sec:magnifications}, lensing reduces the area of
the source plane at high redshifts. Using our cluster lens models, we
estimate the ratio of the source-plane area at $z = 9$ of the lensed
cluster field to the nearby parallel field ranges from $0.24 - 0.85$.
Because of a lack of multiply-imaged sources, the lens models have large
uncertainties, yielding a factor of 3.5 difference in the source-plane
area at high-redshift.

Furthermore, there are more bright foreground galaxies in the cluster
field than the blank field. The galaxy cluster can also hamper
detections somewhat, though advanced methods can model and/or filter
out the brighter cluster light to recover many faint distant galaxies
\citep[e.g.,][]{Livermore2017,Bhatawdekar21}. Using the regions of
segmentation image without source detections, we find that the blank
area of the cluster field is 9\% smaller than the parallel field (3.9
vs. 4.3 arcmin$^2$). Taken together, the high-redshift source plane
area in the cluster field may be $\sim 0.21 - 0.77$ times smaller
than the parallel field. Assuming a similar surface number density
of high-redshift sources over the two fields (although this is not
necessarily the case due to cosmic variance), we would expect reduction
in the number of high-redshift sources in the lensed field by the same
factors.

Our results identify one high-redshift candidate in the cluster field,
\arcid, which with a delensed magnitude of 28.4 (F200W) would have
been detected without lensing. Thus, our number count ratio between
the two fields of 0.33 is consistent with the range of $0.21 - 0.77$.
However, this does not consider the effect of lensing magnification bias
\citep[e.g.,][]{Broadhurst1995}. In a flux-limited sample, the lensed
field, due to the magnification effect, probes galaxies from a fainter
source population than the unlensed field.

At $z \sim 8$, faint number counts in lensed fields should roughly match
those in blank fields \citep[e.g.,][]{Coe15}, given the observed steep
faint-end slope $\alpha \sim -2$ of the UV luminosity function (LF)
\citep[e.g.,][]{Bradley2012b, Bouwens2022}. At higher redshifts $z \ga
9$, we expect steeper LF faint-end slopes, increasing the advantage
for lensing to reveal faint galaxies at these redshifts. More detailed
analyses injecting artificial sources and measuring completeness will be
required of this and other fields to quantify the lensing advantage at
$z \sim 9$ and higher redshifts. Confirmed suppression of lensed number
counts could indicate LF faint-end slopes hovering around $\alpha \sim
-2$ rather than steepening as expected from both simulations and trends
at lower redshifts. However, we can draw no conclusions given the very
small samples sizes presented in this paper.

\subsection{Possible Sources of Contamination}

Low-mass stars, extreme emission-line galaxies (EELG), and photometric
scatter of red low-redshift galaxies can all be sources of contamination
for high-redshift galaxy selections. While low-mass stars and brown
dwarfs can have similar colors as high-redshift galaxies \citep[e.g.,][]
{Yan2003, Ryan2005, Wilkins2014}, we can rule out the possibility
of contamination from these sources because all of our candidates
are resolved (see \S\ref{sec:sizes}). Galaxies with extremely strong
emission lines can sometimes also mimic high-redshift galaxies,
especially in cases with fewer filters longward of the spectral break
\citep{Atek2011, vanderWel2011, Brammer2013}. With our multiband
dataset, we have six filters redward of the break and our SED models
are flexible to include extreme emission lines (as in the case of our
\beagle\ fit to \idnum{8004}, which implies \OIII+\Hb\ rest-frame
equivalent width $\sim$2000 \AA).

Low-redshift red and dust-obscured galaxies can also be a source of
contamination, where their strong Balmer breaks can be confused with
a \Lya\ break. Using \HST\ data, \cite{Bouwens2011} found that the
contamination rates of low-redshift interlopers can be up to 40\%.
However, this can be somewhat mitigated by having many filters redward
of the break. Red or dusty low-redshift interlopers are expected to
have red colors longward of break, which is not the case for three of
our candidates. Three of our candidates have very blue SED longward
of the break, with continuum slopes of $\beta < -2$. The remaining
candidate, \idnum{8004}, has a redder continuum slope of $\beta =
-1.5$. The SED fitting results differ significantly, fitting the redder
continuum either very strong emission lines or a Balmer break, leading
to completely different interpetations for this galaxy. However,
SED fits constrained to low redshifts ($z < 7$) provide much poorer
solutions for this galaxy.

As part of our sample selection, we ran addition SED fits in which we
force the photometric redshifts to $z < 7$ to determine if the breaks
(and fluxes in all bands) can be fit well by a low-redshift solution.
Our selection criteria require a conservative $\chi^2$ difference of at
least 9 between the low-redshift and high-redshift solutions, indicating
much poorer fits for the low-redshift solutions and ruling them out at
$\ge 3\sigma$ significance \citep{Harikane2023}.

We also consider the possibility that we may have detected more
high-redshift candidates in the parallel field because it lacks \HST\
coverage in bluer filters, which might have helped rule out low-redshift
interlopers. To explore this possibility, we re-ran \eazypy\ on the
sources in the lensed field using only their fluxes in the \JWST\ NIRCam
filters, excluding the photometry in all \HST\ filters. After performing
the identical selection criteria as the parallel field, we find only
one additional high-redshift candidate. This would suggest that our
contamination rate is not significantly affected due to the lack of
\HST\ photometry in the parallel field. Formally, inclusion of the \HST\
data excludes 50\% of our candidates in the lensed field, but this
represents only one galaxy. Strong conclusions cannot be drawn from this
small sample size.

While our high-redshift sample selection criteria are designed
to minimize contamination from low-redshift interlopers, the
possibility remains that some of our candidates are at lower redshifts.
Spectroscopic confirmation, like that obtained for \arcid, is needed to
definitively measure the redshifts of these candidates.

\section{Conclusions} \label{sec:conclusions}

We present a search for very high-redshift galaxies in the first \JWST\
NIRCam observations of the lensing cluster \WHL\ and a nearby parallel
field centered $\sim$~2\farcm9 from the cluster center. Combined with
RELICS \HST\ observations, the \JWST\ and \HST\ observations of \WHL\
include imaging in 17 filters spanning $0.4 - 5.0~\micron$ in which we
use to search for high redshift galaxies.

Our final sample of high-redshift candidate galaxies consists of four
candidates with photometric redshifts $\zphot \sim 8.3 - 10.2$. Note
that while our initial \eazypy\ selection was restricted to $\zphot \ge
8.5$, the \bagpipes\ and \beagle\ photometric redshift for one of the
candidates (\idnum{10060}) is at $\zphot = 8.3-8.4$. One $\zphot \sim 9$
candidate, \arcid, lies in the cluster field, while the remaining three
are located in the nearby parallel field.

One candidate, \arcid, is magnified to AB mag 26.2 and stretched into an
arc 2\farcs4 long by the effects of strong gravitational lensing. The
\JWST\ data reveal at least two bright knots of unevenly distributed
star formation within the arc. This candidate also has a magnification
$\mu = 7.9_{-6}^{+12}$ as determined from four independent lens models
of the \WHL\ galaxy cluster. \arcid\ is the most spatially-resolved
galaxy at $\zphot \sim 9$ known to date, similar in length to the
$\zphot \sim 10$ candidate SPT0615-JD1 \citep{Salmon2018}. Recent \JWST\
NIRSpec observations spectroscopically confirm this galaxy at $\zspec =
8.22$ (Vikaeus et al. in prep).

We perform SED fitting to the photometry of these galaxies using the
independent SED-fitting codes \bagpipes\ and \beagle\ to estimate the
physical properties of our candidates. We find stellar masses typically
in the range \logmstar\ = $8.3 - 8.7$, specific star formation rates
sSFR $\sim 0.25 - 10$ Gyr$^{-1}$, young mass-weighted ages $< 240$ Myr,
low dust content with values $A_V < 0.15$, and rest-frame UV slopes of
$\beta = -1.5$ to $-2.6$. We note that the stellar masses and SFR of
the lensed galaxy \arcid\ are highly uncertain because of magnification
uncertainties.

Other \JWST\ analyses have estimated similarly young ages $< 100$
Myr for $z \sim 9 - 16$ candidates \citep{Whitler2023,Furtak2023}.
Discovering such young galaxies is consistent with expectations
from simulations given our image depths down to AB mag $\sim$29
\citep{Mason2023}. Deeper \JWST\ imaging is required to reveal older
(and thus fainter) $\sim$100 Myr populations at $z > 9$, perhaps typical
for the more numerous fainter galaxies in the early universe. Likewise,
\JWST/NIRSpec observations will be needed to spectroscopically confirm
three of these high-redshift galaxy candidates and to study their
physical properties in more detail.


\section{Acknowledgments}

Based on observations with the NASA/ESA/CSA {\em James Webb Space
Telescope} obtained from the Mikulski Archive for Space Telescopes
(MAST) at the Space Telescope Science Institute (STScI), which
is operated by the Association of Universities for Research in
Astronomy (AURA), Incorporated, under NASA contract NAS5-03127.
Support for Program number JWST-GO-02282 was provided through
a grant from the STScI under NASA contract NAS5-03127. The
data described here may be obtained from the MAST archive at
\dataset[doi:10.17909/cqfq-5n80]{https://dx.doi.org/10.17909/cqfq-5n80}.
Also based on observations made with the NASA/ESA {\em Hubble Space
Telescope}, obtained at STScI, which is operated by AURA under NASA
contract NAS5-26555. The \HST\ observations are associated with
programs HST-GO-14096, HST-GO-15842, and HST-GO-16668. Cloud-based data
processing and file storage for this work is provided by the AWS Cloud
Credits for Research program. The Cosmic Dawn Center is funded by the
Danish National Research Foundation (DNRF) under grant \#140. AZ and LF
acknowledge support by Grant No. 2020750 from the United States-Israel
Binational Science Foundation (BSF) and Grant No. 2109066 from the
United States National Science Foundation (NSF), and by the Ministry of
Science \& Technology, Israel.

%

\vspace{5mm}
\facilities{JWST(NIRCam), HST(ACS, WFC3)}


\software{\astropy\ \citep{astropy2022, astropy2018},
          \photutils\ \citep{Bradley2023},
          \grizli\ \citep{Grizli},
          \eazypy\ \citep{Brammer2008},
          \bagpipes\ \citep{bagpipes},
          \beagle\ \citep{beagle},
          \jdaviz\ \citep{jdaviz}
          }



%


\bibliographystyle{aasjournal}



\end{document}